\documentclass[aps,prl,preprint]{revtex4}

\usepackage{epsfig}

\draft

\begin{document}
\author{Hong-Shi Zong$^{1,2}$, Wei-Min Sun$^{1}$, Jia-Lun Ping$^{3}$, Xiao-fu L\"{u}$^{2,4}$, and Fan Wang$^{1}$}
\address {$^{1}$ Department of Physics, Nanjing University, Nanjing 210093, P. R. China} 
\address{$^{2}$ CCAST(World Laboratory), P.O. Box 8730, Beijing 100080, P. R. China}
\address{$^{3}$ Department of Physics, Nanjing Normal University, Nanjing 210097, P. R. China}
\address{$^{4}$ Department of Physics, Sichuan Union University, Chengdu 610064, P. R. China}

\title{Is There Only One Solution of the Dyson-Schwinger Equation for Quark Propagator in the Case of Non-zero Current Quark Mass?}

\begin{abstract}
In this letter it is shown on general ground that there exist two qualitatively distinct solutions of the Dyson-Schwinger equation for the quark propagator in the case of non-zero current quark mass. One solution corresponds to the ``Nambu-Goldstone'' phase and the other one corresponds to the ``Wigner'' phase in the chiral limit.

\bigskip


\bigskip

E-mail: zonghs@chenwang.nju.edu.cn.

\bigskip


\end{abstract}

\maketitle

It is generally believed that with increasing temperature and baryon number density the hadronic matter undergoes a phase transition to the quark-gluon plasma(QGP) which is expected to appear in ultrarelativistic heavy ion collisions. These two phases are generally referred to as the "Nambu-Goldstone" phase(characterized by
dynamical chiral symmetry breaking and confinement of dressed quarks) and the 
"Wigner" phase(corresponding to QGP where chiral symmetry is restored and quarks are not confinement). Theoretically these two phases are described by two different solutions(the "Nambu-Goldstone" and the "Wigner" solution) of the quark propagator and the existence of these two solutions in the chiral limit has been rigorously proved in the framework of Dyson-Schwinger equation(DSE) approach of QCD. However, it is a general view in the literature that when the current quark mass is non-zero, the Dyson-Schwinger equation for the quark propagator has only one solution which corresponds to the "Nambu-Goldstone" phase while the solution  corresponding to the ``Wigner'' phase does not exist[1,2]. But as far as we know, this claim has never been proved. Does there really exist only one solution of the Dyson-Schwinger equation for quark propagator in the case of non-zero current quark mass? 

In order to investigate this problem, let us first recall the usual arguments which exclude the existence of the "Wigner" solution of the Dyson-Schwinger equation for the quark propagator in the case of non-zero current quark mass.  The DSE satisfied by the quark self-energy is
\begin{equation}
\Sigma(p,m)=\int \frac{d^4 q}{(2\pi)^4} g^2_{s}D_{\mu\nu}^{ab}(p-q)\gamma_{\mu}t^a{\cal{G}}(q,m)\Gamma^b_{\nu}(p,q),
\end{equation}
where $g^2_{s}D_{\mu\nu}^{ab}(p)$ and $\Gamma^b_{\nu}(p,q)$ are the full, nonperturbative gluon propagator and the quark-gluon vertex, respectively, $a$, $b$ are colour indices with $t^a=\frac{\lambda^a}{2}$ for the standard Gell-Mann $SU(3)$ representation, and $m$ is the current quark mass.

The quark propagator ${\cal{G}}(p,m)$ and the quark self-energy $\Sigma(p,m)$ are related by
\begin{equation}
{\cal{G}}^{-1}(p,m)\equiv i\gamma\cdot p+m+\Sigma(p,m)\equiv i\gamma\cdot p {\cal{A}}(p^2)+{\cal{B}}(p^2)+m,
\end{equation}
where ${\cal{A}}(p^2)$ and ${\cal{B}}(p^2)$ are the quark self energy functions in the case of non-zero current quark mass. If $m$ is set to be zero, ${\cal{G}}(p,m)$ goes into the dressed quark propagator in the chiral limit $G(p)\equiv{\cal{G}}(p,m=0)$, which has the 
decomposition
\begin{equation}
G^{-1}(p)\equiv i\gamma\cdot p+\Sigma(p,m=0)=i\gamma\cdot p A(p^2)+B(p^2).
\end{equation}

In order to handle Eq.(1) it is necessary to make certain simplifications and truncations. One commonly used truncation of Eq.(1), called the ``rainbow'' approximation, involves replacing the full vertex in Eq.(1) by the bare vertex,
\begin{equation}
\Gamma_{\nu}^{b}(p,q)=t^b\gamma_{\nu}.
\end{equation}
In this case, the quark self energy functions $A(p^2)$ and $B(p^2)$ are
determined by the rainbow DSE in the chiral limit:
\[
[A(p^2)-1]p^2=\frac{4}{3}\int \frac{d^{4}q}{(2\pi)^4}g^2_{s} D(p-q)
\frac{A(q^2)}{q^2A^2(q^2)+B^2(q^2)}\left[p\cdot q+2\frac{p\cdot(p-q)~q\cdot(p-q)}{(p-q)^2}\right],
\]
\begin{equation}
B(p^2)=4\int \frac{d^{4}q}{(2\pi)^4}g^2_{s} D(p-q)
\frac{B(q^2)}{q^2A^2(q^2)+B^2(q^2)},
\end{equation}
where we have used Landau gauge. It is readily seen that $B(p^2)$ in Eq.(5) has two qualitatively distinct solutions. The ``Nambu-Goldstone''
solution, for which $B(p^2)\neq 0$,
describes a phase, in which: 1) chiral symmetry is dynamically
broken, because one has a nonzero quark mass function; and 2) the
dressed quarks are confined, because the propagator described by
these functions does not have a Lehmann representation. The
other solution, the ``Wigner'' one, $B(p^2)\equiv 0$,
describes a phase, in which chiral symmetry is not broken and the dressed-quarks are not confined[2,3]. However, when $m\not= 0$, the self energy functions ${\cal{A}}(p^2)$ and ${\cal{B}}(p^2)$ are
determined by the following rainbow DSE:
\[
[{\cal{A}}(p^2)-1]p^2=\frac{4}{3}\int \frac{d^{4}q}{(2\pi)^4}
\frac{g^2_{s} D(p-q){\cal{A}}(q^2)}{q^2{\cal{A}}^2(q^2)+[{\cal{B}}(q^2)+m]^2}\left[p\cdot q+2\frac{p\cdot(p-q)~q\cdot(p-q)}{(p-q)^2}\right],
\]
\begin{equation}
{\cal{B}}(p^2)=4\int \frac{d^{4}q}{(2\pi)^4}g^2_{s} D(p-q)
\frac{{\cal{B}}(q^2)+m}{q^2{\cal{A}}^2(q^2)+[{\cal{B}}(q^2)+m]^2}.
\end{equation}
Comparing Eq.(6) with Eq.(5), it is apparent that ${\cal{B}}(p^2)\equiv 0$ is not a solution to Eq.(6). From this observation one often concludes that in the case of non-zero current quark mass there exists only one solution(${\cal{B}}(p^2)\not= 0$) for Eq.(6), which corresponds to the Nambu-Goldstone phase in the chiral limit, and the solution  corresponding to the Wigner phase simply does not exist.
However, the fact that ${\cal{B}}(p^2)\equiv 0$ is not a solution to Eq.(6) does not necessarily mean that there exists only one solution(${\cal{B}}(p^2)\not= 0$) for Eq.(6). As will be shown below, there does exist two qualitatively distinct ${\cal{B}}(p^2)\not= 0$ solutions in Eq.(6).

In order to demonstrate this point, let us separate the contributions of dynamical and explicit chiral symmetry breaking(driven by current quark mass $m$) explicitly. By differentiating the dressed quark propagator ${\cal{G}}^{-1}(p,m)$ with respect to $m$, we find that the dressed quark propagator is related to the vertex for the scalar operator $\bar{q}q$ by
\begin{equation}
\Gamma(p,0,m)=\frac{\delta {\cal{G}}^{-1}(p,m)}
{\delta m}.
\end{equation}
Integrating this equation we have 
\begin{equation}
{\cal{G}}^{-1}(p,m)={\cal{G}}^{-1}(p,m=0)+\int^{m}_{0}
\Gamma(p,0,m')dm'\equiv G^{-1}(p)+{\cal{G}}^{-1}_{E}(p,m),
\end{equation}
where $G^{-1}(p)$, an integration constant, is the contribution of dynamical chiral symmetry breaking and is independent of the current quark mass $m$. ${\cal{G}}^{-1}_{E}(p,m)$ is the contribution of explicit chiral symmetry breaking and vanishes if the current quark mass $m$ equals zero. Without loss of generality, ${\cal{G}}^{-1}_{E}(p,m)$ can be written as
\begin{equation}
{\cal{G}}^{-1}_{E}(p,m)=m\left[i\gamma\cdot pE(p^2)+F(p^2)\right].
\end{equation}
Substituting Eqs.(8) and (9) into Eq.(2), we have
\begin{equation}
{\cal{A}}(p^2)\equiv A(p^2)+mE(p^2), ~~~~{\cal{B}}(p^2)\equiv B(p^2)+mF(p^2)-m
\end{equation}
Putting Eq.(10) into Eq.(6), we have the DSE satisfied by $E(p^2)$ and $F(p^2)$
\[
B(p^2)+mF(p^2)-m=4\int \frac{d^{4}q}{(2\pi)^4}g^2_{s} D(p-q)
\frac{B(q^2)+mF(q^2)}{q^2[A(q^2)+mE(q^2)]^2+[B(q^2)+mF(q^2)]^2},
\]
\begin{eqnarray}
[A(p^2)+mE(p^2)-1]p^2&=&\frac{4}{3}\int \frac{d^{4}q}{(2\pi)^4}
\frac{g^2_{s} D(p-q)[A(q^2)+mE(q^2)]}{q^2[A(q^2)+mE(q^2)]^2
+[B(q^2)+mF(q^2)]^2}\nonumber\\
&&\times\left[p\cdot q+2\frac{p\cdot(p-q)~q\cdot(p-q)}{(p-q)^2}\right],
\end{eqnarray}
where $A(p^2)$ and $B(p^2)$ are determined by Eq.(5). 

For a given model gluon propagator $g_s^2D(p)$, we can solve consistently 
Eqs.(5) and (11) to obtain the four scalar functions $A(p^2)$, $B(p^2)$, $E(p^2)$,
 and $F(p^2)$. As was shown above, there exists 
two solutions to Eq.(5), i.e. the ``Nambu-Goldstone'' solution($B(p^2)\neq 0$) and the "Wigner" solution($B(p^2)\equiv 0$). Substituting these two solutions into Eq.(11), we can obtain two different solutions for $E(p^2)$ and $F(p^2)$. For example, substituting $B(p^2)\equiv 0$ into Eqs.(5) and (11), we have the ``Wigner'' solution $A'(p^2)$, $E'(p^2)$ and $F'(p^2)$
\begin{equation}
[A'(p^2)-1]p^2=\frac{4}{3}\int \frac{d^{4}q}{(2\pi)^4}g_{s}^2 D(p-q)
\left[p\cdot q+2\frac{p\cdot(p-q)~q\cdot(p-q)}{(p-q)^2}\right]
\frac{1}{q^2A'(q^2)},
\end{equation}
and
\[
F'(p^2)-1=4\int \frac{d^{4}q}{(2\pi)^4}g^2_{s} D(p-q)
\frac{F'(q^2)}{q^2[A'(q^2)+mE'(q^2)]^2+[mF'(q^2)]^2},
\]
\begin{eqnarray}
[A'(p^2)+mE'(p^2)-1]p^2&=&\frac{4}{3}\int \frac{d^{4}q}{(2\pi)^4}
\frac{g^2_{s} D(p-q)[A'(q^2)+mE'(q^2)]}{q^2[A'(q^2)+mE'(q^2)]^2
+[mF'(q^2)]^2}\nonumber\\
&&\times\left[p\cdot q+2\frac{p\cdot(p-q)~q\cdot(p-q)}{(p-q)^2}\right].
\end{eqnarray}
So far, at the rainbow approximation to the DSE, we have completed the derivation of the dependence of ${\cal{G}}^{-1}(p,m)$ on $m$ in the ``Nambu-Goldstone'' and ``Wigner'' phases separately;
\begin{eqnarray}
{{\cal{G}}}^{(NG)^{-1}}(p,m)=i\gamma\cdot p~A(p^2)+B(p^2)+m\left[i\gamma\cdot p~E(p^2)+F(p^2)\right]~~for~ B(p^2)\not= 0,
\end{eqnarray}
\begin{eqnarray}
{{\cal{G}}}^{(W)^{-1}}(p,m)=i\gamma\cdot p~A'(p^2)+m\left[i\gamma\cdot pE'(p^2)+F'(p^2)\right].
\end{eqnarray}
Just as was shown by Eqs.(14-15), there does exist two qualitatively distinct solutions with ${\cal{B}}(p^2)\not= 0$ in the case of non-zero current quark mass. In addition, we want to stress that the above approach is general in the sense that it does not depend on the rainbow approximation used here. It can also be applied to the case of finite chemical potential $\mu$. By separating the contributions of dynamical and explicit chiral symmetry breaking(driven by chemical potential $\mu$), one can study the chemical potential dependence of the dressed-quark propagator and to find without arbitrariness solutions representing the ``Nambu-Goldstone'' phase and the ``Wigner'' phase at non-zero chemical potential[4]. 

In order to have a qualitative understanding of the above two qualitatively distinct solutions, a particularly simple and useful model of the dressed gluon two-point function[5] is employed:
\begin{equation}
g_{s}^2 D_{\mu\nu}(p-q)=4\pi^4\eta^2\left[\delta_{\mu\nu}-\frac{(p-q)_{\mu}(p-q)_{\nu}}{(p-q)^2}\right]\delta^{(4)}(p-q),
\end{equation}
where the scale parameter $\eta$ is a measure of the strength of the infrared slavery effect. This model has the advantage that the integral equation of DSE reduce to algebraic equations.

Substituting then Eq.(16) into Eqs.(5) and (12), we have the ``Nambu-Goldstone'' solution; 
\begin{eqnarray}
B(p^2)&=&(\eta^2-4 p^2)^{\frac{1}{2}},~~~~~~~A(p^2)=2 ~~~~for~~~~ 
p^2 < \frac{\eta^2}{4},\nonumber\\
B(p^2)&=&0,~~~~A(p^2)=\frac{1}{2}\left[1+(1+\frac{2\eta^2}{p^2})^{\frac{1}{2}}
\right]~~~for~~~~ p^2\geq\frac{\eta^2}{4},
\end{eqnarray}
and the ``Wigner'' solution in chiral limit;
\begin{eqnarray}
B'(p^2)&\equiv&0,~~~~A'(p^2)=\frac{1}{2}\left[1+(1+\frac{2\eta^2}{p^2})^{\frac{1}{2}}
\right].
\end{eqnarray}

With the model of the dressed gluon propagator specified in Eq.(16) and the explicit expression for $A(p^2)$, $B(p^2)$ and $A'(p^2)$ given in Eqs.(17,18), Eqs.(11) and (13) entail that the scalar functions $E(p^2)$, $F(p^2)$ satisfy
\[
A(p^2)+mE(p^2)=\frac{2[B(p^2)+mF(p^2)]}{B(p^2)+mF(p^2)+m},
\]
\begin{equation}
\left[B(p^2)+mF(p^2)-m\right]\left\{\frac{4p^2[B(p^2)+mF(p^2)]}{[B(p^2)+mF(p^2)+m]^2}+[B(p^2)+mF(p^2)]\right\}=\eta^2,
\end{equation}
and the scalar functions $E'(p^2)$, $F'(p^2)$ satisfy 
\[
A'(p^2)+mE'(p^2)=\frac{2F'(p^2)}{F'(p^2)+1},
\]
\begin{equation}
\left[F'(p^2)-1\right]\left\{\frac{4p^2F'(p^2)}{[F'(p^2)+1]^2}+m^2F'(p^2)\right\}=\eta^2.
\end{equation}
Using Mathematica, it is not difficult to verify that the solutions of Eqs.(19) and (20) do exist. The full expressions for $E(p^2)$, $F(p^2)$, $E'(p^2)$, and 
$F'(p^2)$ are too lengthy and will not be given in this letter. It should be noted that the model gluon propagator(16) is an infrared-dominant model that does not represent well the behavior of $g_{s}^2D_{\mu\nu}(p)$ away from $p^2\simeq 0$. Nevertheless, this simple model can provide a reasonable guide to the nonperturbative properties of more sophisticated DSE-model of QCD.

To summarize: we show on general ground that there exist two qualitatively distinct solutions of the Dyson-Schwinger equation for the quark propagator in the case of non-zero current quark mass(one of which corresponds to the ``Nambu-Goldstone'' phase and the other corresponds to the ``Wigner'' phase in the chiral limit). This approach has the advantage that we can analyze the effects of explicit and dynamical chiral symmetry breaking separately. The basic equations used here are Eqs.(5) and (11). In order to have a qualitative understanding of the above two qualitatively distinct solutions, we choose a simple, confining model(16) in solving Eqs.(5) and (11). From this the dressed quark propagator in the ``Nambu-Goldstone'' phase and the ``Wigner'' phase are derived. With these two ``phases'' characterized by qualitatively different momentum-dependent quark propagators, one can study the QCD phase structure in a definite way. 

\vspace*{0.5 cm}
\noindent{\large \bf Acknowledgments}

This work was supported in part by the National Natural Science Foundation of China(under Grant Nos 10175033, 10135030) and the Research Fund for the Doctoral Program of Higher Education(under Grant No 20030284009).

\vspace*{1.0 cm}
\noindent{\large \bf References}

\begin{description}

\item{[1]} R. T. Cahill, C. D. Roberts, Phys. Rev. {\bf D32}, 2419 (1985).
\item{[2]} P. C. Tandy, Prog. Part. Nucl. Phys. 39, 117 (1997), and references therein.
\item{[3]} C. D. Roberts and A. G. Williams, Prog. Part. Nucl. Phys. {\bf 33}, 477 (1994), and references therein.
\item{[4]} Hong-shi Zong, et.al., in preparation. 
\item{[5]} H. J. Munczek and A. M. Nemirovsky, Phys. Rev. {\bf D28}, 181 (1983).

\end{description}
\end{document}